\documentclass[notitlepage,twocolumn,prl,longbibliography]{revtex4-1}
\usepackage{times}
\usepackage{graphicx}
\usepackage{bm}
\usepackage{amssymb,amsfonts,amsmath,amsbsy,bm,t1enc,latexsym}
\usepackage{soul}
\usepackage[super]{nth}
\usepackage{hyperref}
\usepackage{setspace}

\begin{document}	
	
	\title{Massively parallel coherent laser ranging using soliton microcombs}
	
	\author{Johann Riemensberger}
	\author{Anton Lukashchuk}
	\author{Maxim Karpov}
	\author{Wenle Weng}
    \author{Erwan Lucas}
    \author{Junqiu Liu}
	\author{Tobias J. Kippenberg}
	\email{tobias.kippenberg@epfl.ch}
	\affiliation{{\'E}cole Polytechnique F{\'e}d{\'e}rale de Lausanne (EPFL), Laboratory of Photonics and Quantum Measurements (LPQM), Lausanne, CH-1015, Switzerland}
	
	\date{\today}
	
	\maketitle


\textbf{
Coherent ranging, also known as frequency-modulated continuous-wave (FMCW) laser based ranging (LIDAR)~\cite{Bostick1967} is currently developed for long range 3D distance and velocimetry in autonomous driving~\cite{Urmson2008,Behroozpour2017}. 
Its principle is based on mapping distance to frequency~\cite{MacDonald1981,Uttam1985}, and to simultaneously measure the Doppler shift of reflected light using frequency chirped signals, similar to Sonar or Radar~\cite{Gnanalingam1952,Hymans1960}. 
Yet, despite these advantages, coherent ranging exhibits lower acquisition speed and requires precisely chirped~\cite{Roos2009} and highly-coherent \cite{Uttam1985} laser sources, hindering their widespread use and impeding Parallelization, compared to modern time-of-flight (TOF) ranging that use arrays of individual lasers. 
Here we demonstrate a novel massively parallel coherent LIDAR scheme using a photonic chip-based microcomb~\cite{Kippenberg2018}. 
By fast chirping the pump laser in the soliton existence range~\cite{Lucas2017a} of a microcomb with amplitudes up to several GHz and sweep rate up to 10 MHz, the soliton pulse stream acquires a rapid change in the underlying carrier waveform, while retaining its pulse-to-pulse repetition rate. 
As a result, the chirp from a single narrow-linewidth pump laser is simultaneously transferred to all spectral comb teeth of the soliton at once, and allows for true parallelism in FMCW LIDAR. 
We demonstrate this approach by generating 30 distinct channels, demonstrating both parallel distance and velocity measurements at an equivalent rate of 3~Mpixel/s, with potential to improve sampling rates beyond 150~Mpixel/s and increase the image refresh rate of FMCW LIDAR up to two orders of magnitude without deterioration of eye safety.
The present approach, when combined with photonic phase arrays~\cite{Mcmanamon1996} based on nanophotonic gratings~\cite{Sun2013}, provides a technological basis for compact, massively parallel and ultra-high frame rate coherent LIDAR systems.}
\noindent In recent years there has been a major interest in LIDAR fueled by the emergent development of autonomous driving~\cite{Urmson2008}, which require the ability to quickly recognize and classify objects in fast-changing and low visibility conditions~\cite{Levinson2011}. 
LIDAR can overcome challenges of camera imaging, such as those associated with weather conditions or illumination, and was used successfully in nearly all recent demonstrations of high-level autonomous driving~\cite{Maddern2017}. 
Generally, laser ranging is based on two different principles; time-of-flight (TOF) and coherent ranging \cite{Bosch2001}. 
In TOF LIDAR, the distance of an object is determined based on the delay of reflected laser pulses. 
To increase the speed of image acquisition, modern systems employ an array of individual lasers (as many as 256) to replace slow mechanical scanning~\cite{Schwarz2010}. 
The velocity information can only be inferred by comparing subsequent images, which is prone to errors due to vehicle motion and interference. 

A different principle is that of frequency modulation continuous-wave (FMCW) LIDAR~\cite{Bostick1967,MacDonald1981,Uttam1985}. 
In this case a laser that is linearly chirped is sent to an object, and the time-frequency information of the return signal is determined by delayed homodyne detection. 
The maximum range is therefore limited not only by the available laser power but also the coherence length of the laser \cite{Uttam1985}.
Assuming a triangular laser scan (over an excursion bandwidth $B$ with period $T$, cf.~Fig.~1e), the distance information (i.e. time-of-flight $\Delta t$) is mapped to a beat note frequency ~\cite{MacDonald1981}, i.e. $\bar{f}=\Delta t \cdot 2B/T$ for a static object. 
Due to the relative velocity $(v)$ of an object, the returning laser light is detected with a Doppler shift $\Delta f_D = \overrightarrow{k}\cdot \overrightarrow{v}/\pi$, where $\overrightarrow{k}$ is the wavevector and $\overrightarrow{v}$ the velocity of the illuminated object.
As a result, the homodyne return signal for a moving object is composed of two frequencies for the upwards and downwards laser scan, i.e. $f_u=\bar{f} + \Delta f_D $ and $f_d = \vert -\bar{f} + \Delta f_D \vert$. 
From the measured beat notes during one period of the scan, one can therefore determine both distance and relative velocity of an object (cf.~Fig.~\ref{fig_concept}e). 
The latter greatly facilitates image processing and object classification, particularly relevant to traffic. 
Moreover, FMCW LIDAR increases the photon flux used for ranging, hence increasing sensitivity and range compared to time-of-flight LIDAR systems, which to date rely on sequential switching of laser diode arrays. 
Furthermore, coherent LIDAR is superior to time-of-flight implementations in low visibility and high background light conditions, culminating in achievements such as ranging objects engulfed in flames~\cite{Mitchell2018}, as delayed homodyne detection makes it almost impervious to interference and malicious remote attacks~\cite{Petit2015}.
Despite these advantages, coherent ranging suffers from the stringent requirement of narrow linewidth \cite{Uttam1985}, as well as fast and linear frequency chirping \cite{Roos2009}, which makes massively parallel implementations, as used in time-of-flight LIDAR, challenging.


\subsection*{Concept of soliton-based parallel FMCW ranging}

\begin{figure*}[!htbp]
	\includegraphics[width=\linewidth]{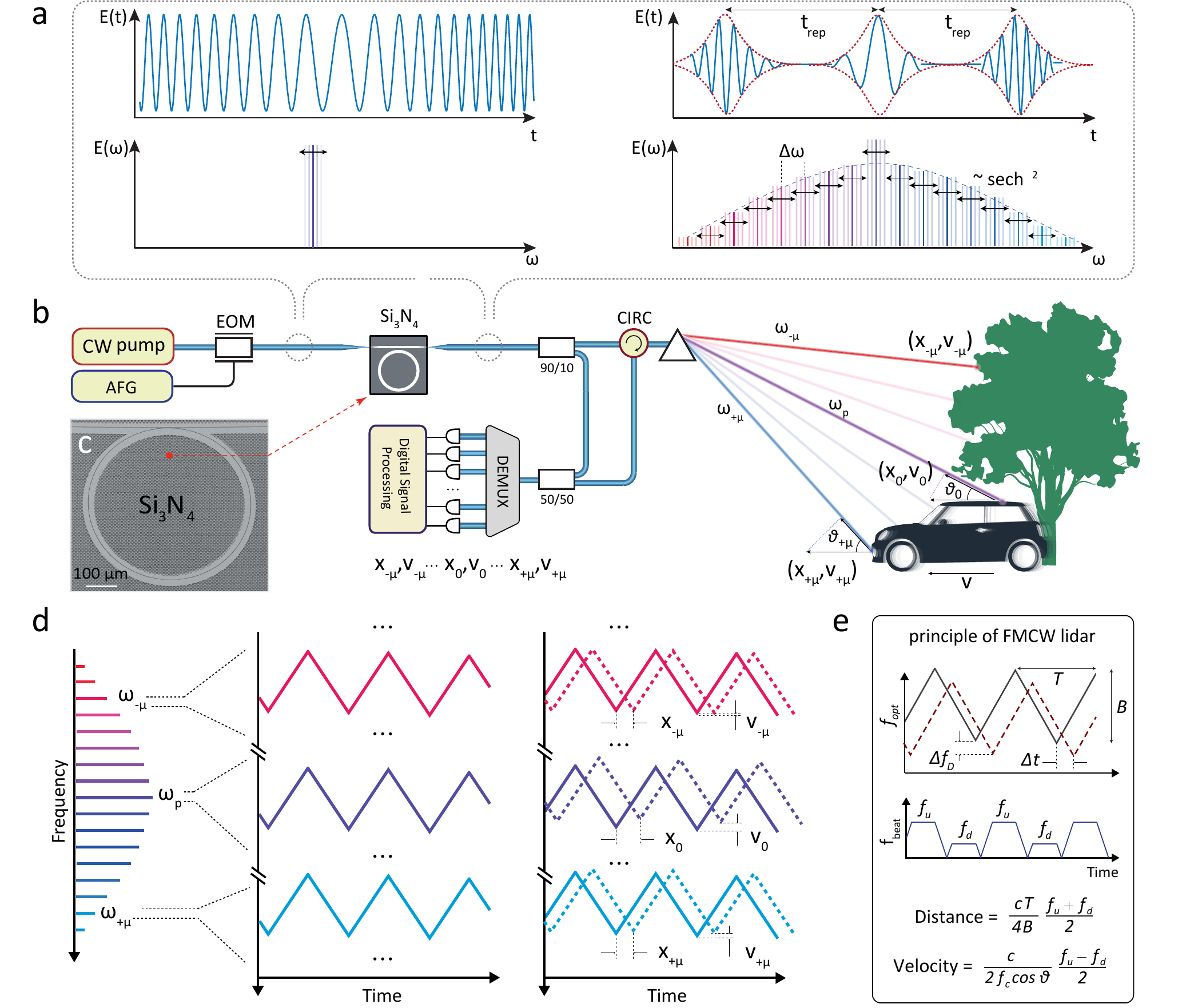}
	\caption{\footnotesize  \linespread{1}  \textbf{Massively parallel frequency-modulated continuous-wave LIDAR using soliton microcombs.} 
		a)~Principle of dissipative Kerr soliton (DKS) generation in a microresonator with a frequency agile laser. By chirping the continuous wave pump laser, the soliton pulse stream exhibits a change in the underlying carrier, while the pulse to pulse repetition rate remains unchanged. In the frequency domain this corresponds to scanning each individual comb tooth, i.e. a change of the carrier envelope frequency only. 
		b)~Schematic outline of the proposed system design. A frequency modulated pump laser drives a photonic integrated $\mathrm{Si_3N_4}$ microresonator. Each individual sideband, spatially dispersed with diffractive optics, serves as source of FM modulated laser light in a parallel detection scheme.
		c)~Electron microscope picture of 228.43 $\mu$m $\mathrm{Si_3N_4}$ microring resonator.
		d)~Principle of coherent velocimetry and ranging with multiple optical carriers isolated from a soliton microcomb. Interleaved upwards and downwards frequency slopes map the distance and radial velocity of target objects onto the mean and the separation of two intermediate frequency beat tones in a delayed homodyne detection scheme. 
		e)~Schematic of the detected beat notes arising in coherent LIDAR ranging of a moving object. The reflected laser light is both time delayed, and frequency shifted due to the Doppler effect, leading to the observation of two homodyne beat notes during one scan period of the laser.}
	\label{fig_concept}
\end{figure*} 

Here we demonstrate a massively parallel coherent FMCW source based on a soliton microcomb integrated on a photonic chip. 
Specifically, we show that agile chirping of the pump laser $\omega_p$ retains the soliton state and leads to simultaneous chirping of all comb teeth $\omega_{\pm \mu}$ comprising the soliton. 
The principles of massively parallel coherent LIDAR based on soliton microcombs are illustrated in Fig.~\ref{fig_concept}a. The underlying idea is to transfer the chirp of a prepared FM LIDAR source to multiple comb sidebands by using it to generate a dissipative Kerr soliton (DKS) \cite{Herr2014b,Leo2010}. 
In the time domain (cf.~Fig.~\ref{fig_concept}a), we modulate the underlying soliton carrier frequency, while minimizing changes of the pulse envelope and repetition rate. 
In the frequency domain, this corresponds to a concurrent modulation of the optical frequency of each comb tooth around its average value (i.e. a modulation of the frequency comb's carrier-envelope frequency). 
This effect, when combined with triangular frequency modulation of a narrow linewidth pump laser, generates a massively parallel array of independent FMCW lasers. 
When dispersing the channels using diffractive optics, as illustrated in Fig.~\ref{fig_concept}b, each channel can acquire both distance and velocity information \emph{simultaneously} (cf.~Fig.~\ref{fig_concept}d). 

The novel scheme leverages three key properties of DKS; the large (i.e. GHz) existence range of the soliton, the fact that repetition rate changes associated with laser scanning are small, and the possibility, as detailed below, to very rapidly sweep between stable operating points without destroying the soliton state or deterioration of the chirp linearity. 
Homodyning the reflected signal with the original comb teeth channel-by-channel, using low bandwidth detectors and digitizers, allows the coherent ranging signal to be recovered and reconstructed for \emph{each comb line $ \mu $ simultaneously}, yielding velocity and distance $(x_\mu, v_\mu)$ for each pixel. 
The presented scheme thus enables true parallel detection of dozens and potentially hundreds of pixels simultaneously. 
Hence, massively parallel - and high speed - coherent LIDAR becomes possible, while requiring only a single well-controlled laser to generate the carrier-frequency chirped soliton. 
This contrasts our approach from dual-frequency comb coherent time-of-flight systems~\cite{Suh2018a,Trocha2018}, which on the other hand achieve best distance precision and acquisition speeds, yet exhibit a limited ambiguity range dictated by the pulse repetition rate, and are challenging to parallelize as the whole frequency comb must illuminate a single pixel. 
In a similar fashion, recently demonstrated coherent stitching of multiple channels from an electro-optical frequency comb generator can be used to improve the distance measurement accuracy of FMCW~\cite{Kuse2019}, yet demands the spectral overlap of adjacent comb modes and concurrent illumination of a single pixel.


\begin{figure*}[!htbp]
	\includegraphics[width=\linewidth]{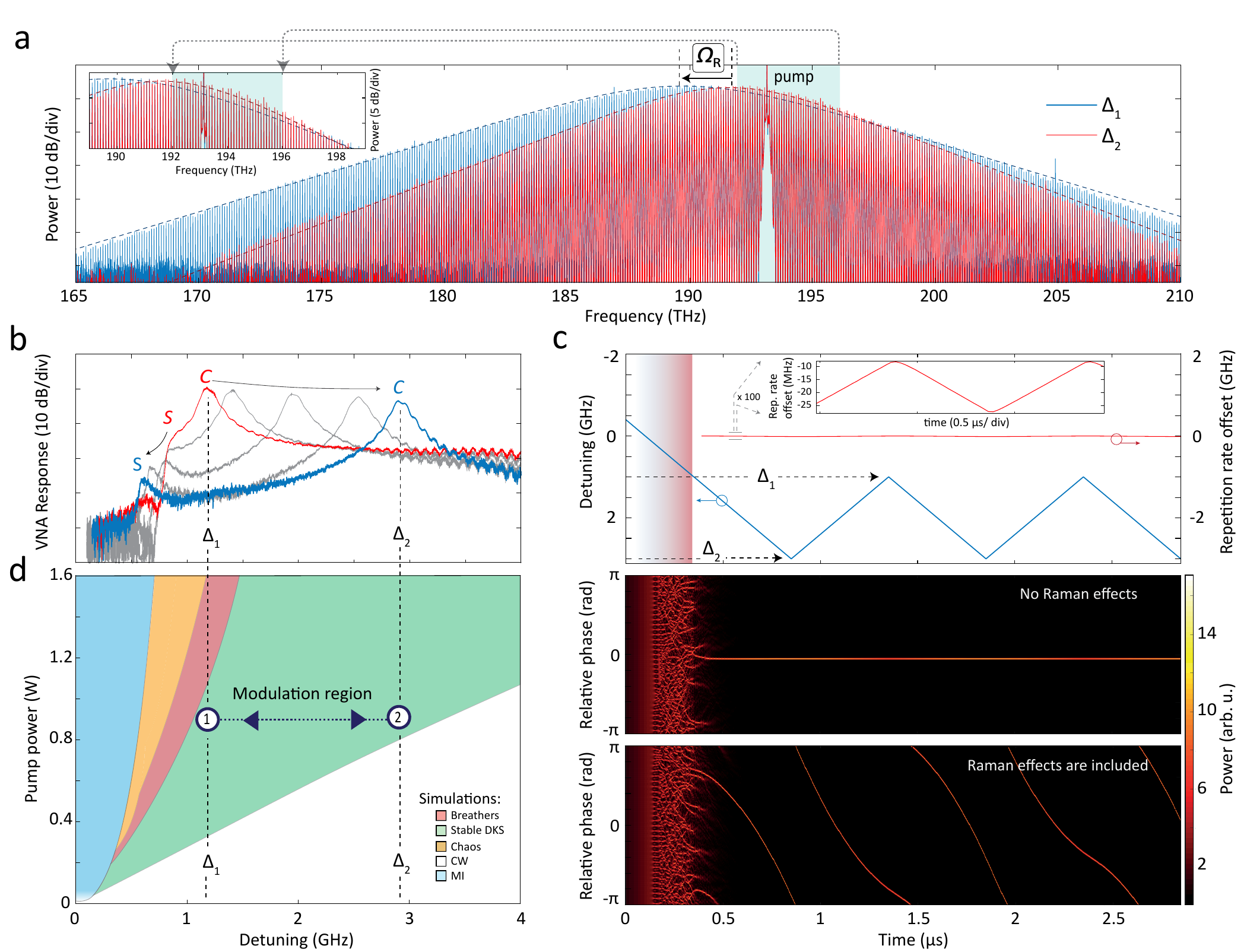}
	\caption{\footnotesize  \linespread{1}  \textbf{Dynamics of frequency-modulated soliton microcombs} 
		a)~Optical spectra of a dissipative Kerr soliton (DKS) at relative laser cavity detuning $\Delta_1=$1.2~GHz and $\Delta_2=$2.9~GHz, respectively. The Raman self-frequency shift $\Omega_\mathrm{R}/2\pi$ of the soliton is highlighted. Inset: Spectral region of FM LIDAR operation showcasing individual line flatness better than 3 dB over the full pump laser frequency excursion range. 
		b)~Phase modulation response of DKS measured with network analyzer (VNA). 
		c)~Numerical simulation results of FM DKS. The laser is tuned through the modulation instability region (blue shaded area) and the breathing region (red shaded area) into the soliton state and triangular FM is imprinted at a chirp rate of $(d\omega_L/dt)/2\pi\approx$ $1.7\cdot10^{16}$~Hz$^2$. The inclusion of stimulated Raman scattering into the simulation reveals a modulation of the repetition rate of up to 10~MHz during the FM cycle. 
		d)~Simulated stability chart of the soliton microcomb for the device used in the LIDAR experiments. The soliton existence range is highlighted in green and confined by the stability of the soliton solution.}
	\label{fig_numerical_simulations}
\end{figure*}

We demonstrate the principle of spectral multiplexing in coherent LIDAR employing a 99~GHz repetition rate DKS in a silicon nitride ($\mathrm{Si_3N_4}$) microresonator, which is fabricated using the photonic Damascene process~\cite{Pfeiffer2016} (cf.~Inset of Fig.~\ref{fig_concept}b and methods section).
Fig.~\ref{fig_numerical_simulations}a) shows the optical spectra of the DKS at the extremal points of the soliton existence range with relative laser-cavity detuning $\Delta_1 = 1.2$~GHz and $\Delta_2 = 2.9$~GHz.
Increasing the detuning, we observe well known temporal compression (58~fs to 45~fs) \cite{Lucas2017a} and Raman self-frequency shift ($\Omega_\mathrm{R}/2\pi = 2$~THz) \cite{Karpov2016} of the DKS. 
Interestingly, despite the frequency excursion strongly exceeding the overcoupled cavity linewidth ($\kappa_0/2\pi = 15$~MHz, $\kappa_{\mathrm{ex}}/2\pi = 80$~MHz) the power of comb teeth between 190~THz and 200~THz does not change by more than 3~dB, providing therefore more than 90 channels suitable for coherent LIDAR. 
The relative laser detuning can be inferred from the phase modulation response spectrum (cf.~Fig.~\ref{fig_numerical_simulations}b), wherein the $C$-resonance peak directly reveals the relative detuning between the cavity resonance and the CW pump laser \cite{Guo2017}.

We next perform numerical simulations based on the Lugiato-Lefever equation~(LLE)~\cite{Lugiato1987,Chembo2013}, which demonstrate the ability of the DKS state to transfer the chirp from the pump to all comb teeth (cf.~Fig.~\ref{fig_numerical_simulations}c). 
The numerical laser scan is started at $\Delta=-0.4$~GHz and the detuning is subsequently increased with a linear chirp rate of $\vert \frac{d \Delta}{dt} \vert = 10^{16}$~Hz$^2$, tuning past the modulation instability region (MI) exciting a single soliton. 
Hereafter, the linear laser scan is inflected and a symmetric triangular FM with equal chirp rate is continued. 
If stimulated Raman effects~\cite{Karpov2016,Yi2017} and higher order dispersion are neglected, the repetition rate remains almost perfectly constant and the frequency chirp is faithfully transduced to each comb line.
Even more surprisingly, the inclusion of stimulated Raman scattering and third order dispersion effects, only induces a small repetition rate mismatch $\Delta f_\mathrm{rep}$ of 20.6~MHz per 1.7~GHz of laser tuning, which is observed as acceleration and deceleration of the soliton in the cavity (cf.~Fig.~\ref{fig_numerical_simulations}d,~bottom).

The linear dependence $\frac{d f_\mathrm{rep}}{d\Delta} \approx \frac{\Omega_\mathrm{R}}{2\pi} \frac{D_2}{D_1} $ results in a spectral channel-dependent bandwidth and, hence, a constant rescaling factor of the measured LIDAR distance, which we can determine during calibration. 
Only nonlinear dependencies of the pulse repetition rate $f_{\mathrm{rep}}$ on the detuning $\Delta$, from either the Raman shift \cite{Yi2016} or multimodal interactions \cite{Yi2017} actually degrades the linearity of the transduced chirp. 
The maximum detuning, which still supports stable DKS generation is determined by the input pump power~\cite{Lucas2017a}, which in term is fundamentally limited by a Raman instability~\cite{Wang2018}.



\begin{figure*}[!htbp]
	\includegraphics[width=\linewidth]{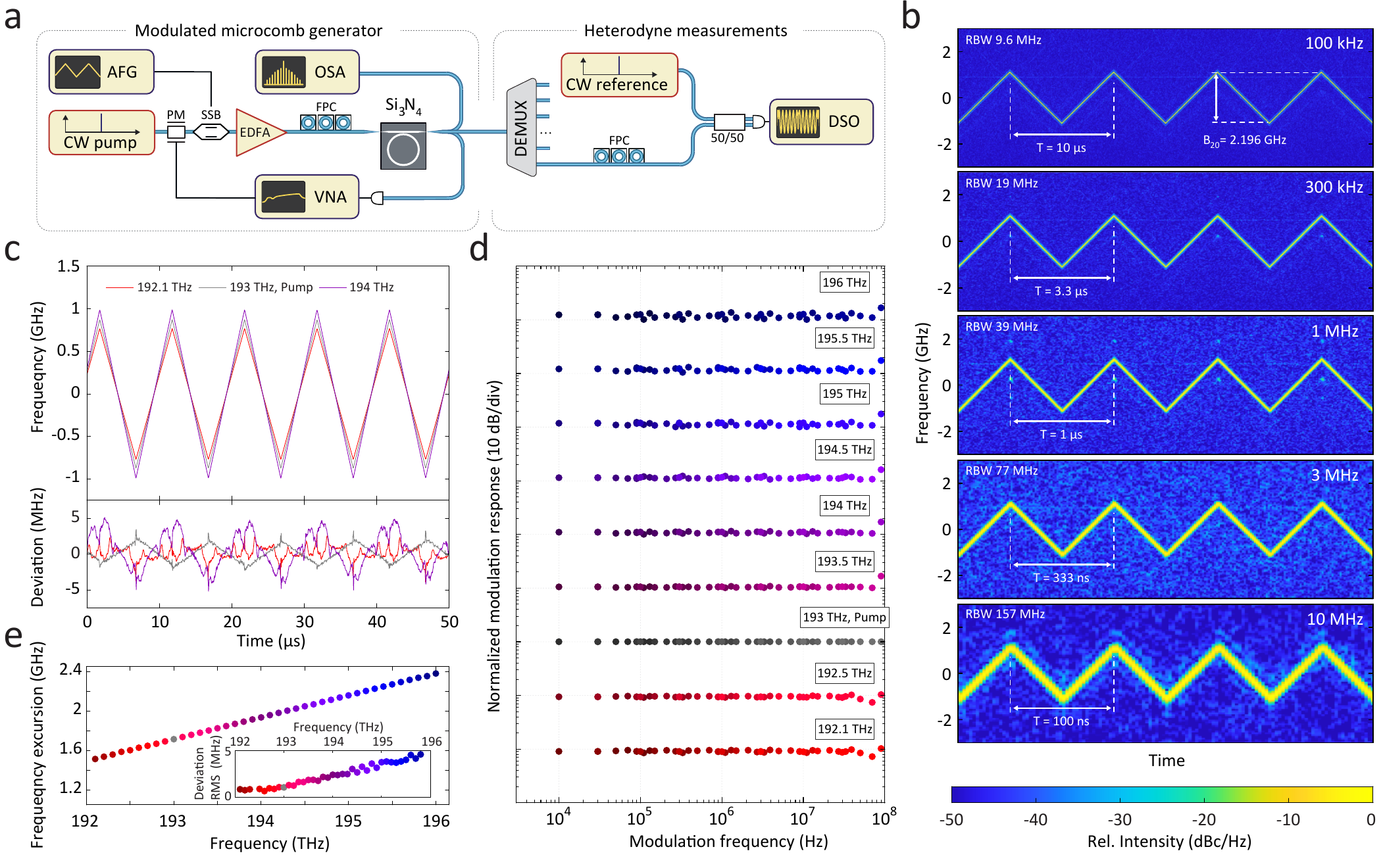}
	\caption{\footnotesize  \linespread{1} \textbf{Time-frequency analysis of a chirped soliton microcomb.} 
		a)~Experimental setup. An amplified external cavity diode laser (ECDL) laser at 193~THz generates a soliton microcomb on the photonic chip. Frequency modulation is applied with a single sideband modulator (SSB). The time-dependent sideband frequencies are detected by beating with a second tuneable ECDL. Optical spectrum analyzer (OSA) and vector network analyzer (VNA) are used for soliton state characterization (cf. Fig.\ref{fig_numerical_simulations}), only.
		b)~Time-frequency maps of 1.6~GHz pump laser chirps at modulation frequencies from 10~kHz to 10~MHz, detected at the $\mu=+20$ comb sideband~(195~THz). 
		c)~Instantaneous frequency of the heterodyne beat note (top) determined by short-time Fourier transform. Deviation from a perfect triangular scan calculated by least-squares fitting (bottom) at modulation frequency 100~kHz.
		d)~Pump to sideband FM transduction determined from the FM amplitude of the first~9~harmonics of the modulation frequency of linearized FM traces between 10~kHz~(dark shading) and 10~MHz~(light shading). Sideband values (colors same as panel~e) are offset by 10~dB and normalized with respect to the modulation amplitudes of the pump. 
		e)~Channel-dependent frequency excursion bandwidth at 100~kHz modulation frequency. Inset: RMS of deviation from perfectly triangular modulation pattern.} 
	\label{fig_chirplinearity_and_linewidth}
\end{figure*}

\subsection*{Characterization of parallel FMCW LIDAR source}
\noindent Next, we experimentally demonstrate he ability to faithfully transfer the pump laser chirp to the soliton microcomb sidebands (cf.~Fig.~\ref{fig_chirplinearity_and_linewidth}).
Details of the experimental setup for heterodyne characterization, linearization of the triangular frequency modulation patterns, and transduction data analysis are described in the methods section. 
Results for the comb tooth at 195~THz ($\mu=+20$) and modulation frequencies $1/T$ from 100~kHz to 10~MHz are depicted in Fig.~\ref{fig_chirplinearity_and_linewidth}b and in the extended data Fig.~\ref{fig_heterodyne_allchannels}. 
The frequency excursion bandwidth $B_{\mu}$ increases linearly with the channel number $\mu$ (cf.~Fig.~\ref{fig_chirplinearity_and_linewidth}e) at a rate of $\frac{d B_{\mu}}{d\mu} = 22.15$~MHz in agreement with the predictions from numerical simulations including stimulated Raman scattering (cf.~Fig.~\ref{fig_numerical_simulations}c). 
We define the chirp nonlinearity as the deviation of the measured instantaneous frequency from a perfectly symmetric triangular FM scan, estimated with least-squares fitting, and depict results for the pump and two comb teeth in Fig.~\ref{fig_chirplinearity_and_linewidth}c (lower panel). Narrow peaks of the chirp nonlinearity are attributed to single-mode dispersive waves~\cite{Yi2017}. We do not observe intermode breathing of the soliton~\cite{Guo2017a} in the present system. The channel dependent RMS nonlinearity is depicted in the inset of Fig.~\ref{fig_chirplinearity_and_linewidth}e and remains below 1/500 of the full frequency excursion for all channels at 100 kHz modulation frequency. 
The frequency-dependent FM transduction is from the pump laser to the DKS teeth calculated from the transduced chirps (cf. extended data Fig.~\ref{fig_heterodyne_analysis_method}) and is plotted in Fig.~\ref{fig_chirplinearity_and_linewidth}d.
We find a lower bound for the 3~dB modulation frequency cutoff of 40~MHz, which corresponds to a maximum per-channel chirp rate of~$1.6\cdot10^{17}$~Hz$^2$. The estimated accumulated chirp rate of all channels thus rivals state-of-the-art swept source lasers, which achieve chirp rates of $10^{18}-10^{19}$~Hz$^2$~\cite{Klein2013}. 


\begin{figure*}[!htbp]
	\includegraphics[width=\linewidth]{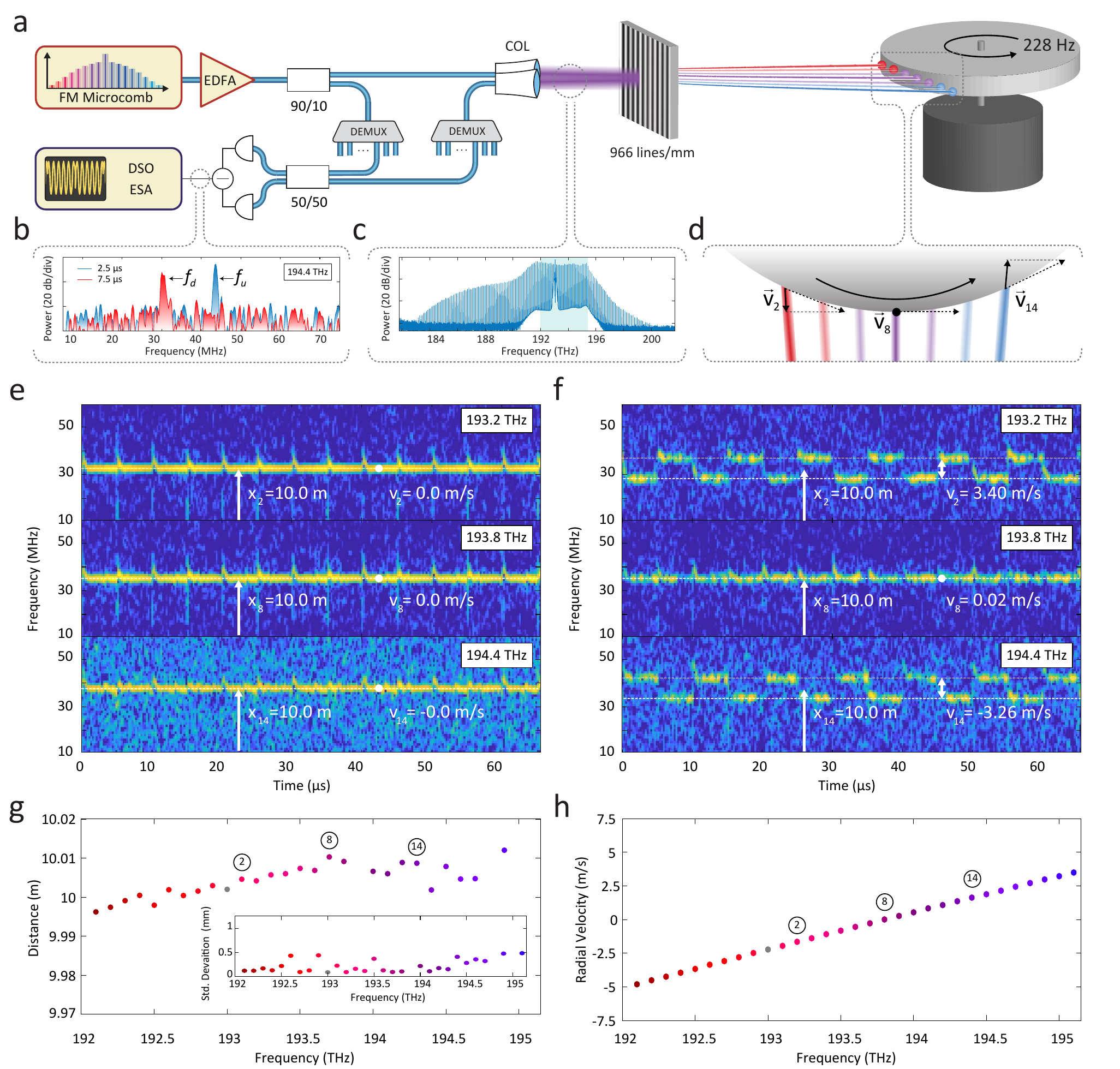}
	\caption{\footnotesize \linespread{1} \textbf{Demonstration of massively parallel velocity measurement using a soliton microcomb.} 
		a)~Experimental setup. The amplified FM LIDAR microcomb source is split into signal and local oscillator (LO) pathways. The signal is dispersed with a transmission grating (966~lines/mm) over the horizontal circumference of a flywheel mounted on a small DC motor. The reflected signals are spectrally isolated before detection.
		b)~RF spectrum of LIDAR backreflection mixed with the LO (sampling length 3.75~$\mu$s) around 2.5~$\mu$s (upward ramp) and 7.5~$\mu$s (downward ramp).
		c)~Optical spectrum of comb lines after amplification. Blue shading highlights 30 comb lines with sufficient power ($> 0$~dBm) for LIDAR detection.s
		d)~Schematic illustration of the flywheel section irradiated by the FM soliton microcomb lines indicating the projection of the angular velocity of the wheel onto the comb lines. 
		e)~Time-frequency maps of selected microcomb FMCW LIDAR channels (sampling length 0.5~$\mu$s) for the static flywheel. 
		f)~Same as e) but for flywheel rotating at 228~Hz.
		g)~Multichannel distance measurement results for the static flywheel. Distance measurement not corrected for fiber path difference between signal and LO path.
		h)~Multichannel velocity measurement for the flywheel rotating at 228~Hz. The accuracy of distance and velocity measurements in case of the rotating flywheel is limited by vibrations.}
	\label{fig_lidar_flywheel}
\end{figure*}

\subsection*{Parallel ranging, velocimetry and 3D imaging}
\noindent Next, we perform a proof-of-concept demonstration of the massively parallel LIDAR system. 
The calibrated FM microcomb is split (90/10) into a signal path, which is spectrally dispersed around the circumference of the flywheel by a transmission grating (966 lines/mm), and a local oscillator path. The spectral channels of the reflected signal and the LO are isolated using a bidirectional arrayed waveguide grating.
The results of parallel distance and velocity measurement including standard deviations over 100~FM periods for the static wheel are displayed in Fig.~\ref{fig_lidar_flywheel}e,g. 
Channels beyond 195.2~THz are not observed with sufficient signal-to-noise ratio~(SNR), because of limited amplification bandwidth.
The measurement imprecision over 25~spectral channels is below 1~cm, comparable with state-of-the-art TOF LIDAR systems and can be improved by using more broadband chirps. 
Small systematic offsets on the level of 1~cm are associated to the lengths of fibre pigtails in the demultiplexers and switches.
The results for the wheel spinning at 228~Hz are depicted in Fig.~\ref{fig_lidar_flywheel}f, resolving the position dependency of the projected velocity around the circumference of the wheel (cf. Fig.~\ref{fig_lidar_flywheel}d). The measurement accuracy in case of the spinning wheel is limited by vibration.
The equivalent distance and velocity sampling rate of the 30 independent channels is 3~Mpixel/s.


\begin{figure*}[!htbp]
	\includegraphics[width=\linewidth]{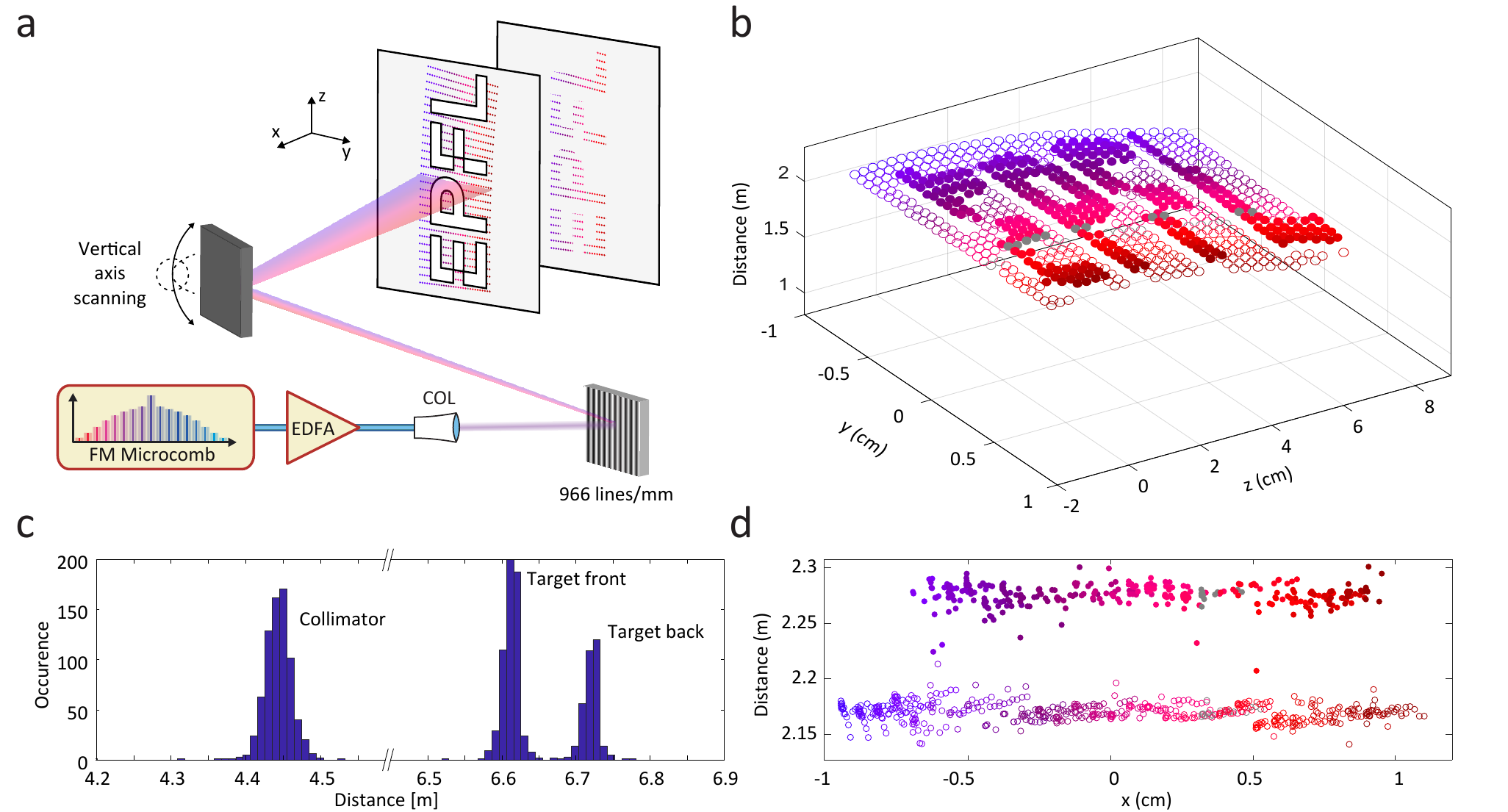}
	\caption{\footnotesize  \linespread{1} \textbf{Parallel distance measurement and imaging} 
		a)~Experimental setup. 
		30 channels of the soliton microcomb are spectrally dispersed with a transmission grating in the horizontal axis (y). 
		Vertical translation is performed by a planar mirror placed behind the grating.  
		The target is formed by two vertical sheets of paper placed at a distance of 11.5~cm. 
		The EPFL logo is cut out from the first sheet. 
		The colored dots mark the approximate positions of the individual beams during the scan and denote the individual spectral channels according to Fig.~\ref{fig_chirplinearity_and_linewidth}e.
		b)~Obtained 3D image by scanning the beam array in the vertical direction. 
		Filled circles denote pixels detected in the target back plane.
		c)~Histogram of successful detections for the collimator (zero distance plane), front and back planes. 
		d)~Projection of b) along the z-axis reveals the cm level distance measurement accuracy and precision for the 30 FM LIDAR channels.}
	\label{fig_lidar_logo}
\end{figure*}

Last, we demonstrate parallel 3D imaging of 30 channels spectrally dispersed with a transmission grating and concurrently illuminating a target composed of two sheets of white paper spaced by 11~cm with the "EPFL" logo cutout in the front plane (cf.~Fig.~\ref{fig_lidar_logo}). 
The target profile is imaged by translation of the beams in the vertical direction with a $45^{\circ}$ steering mirror, and depicted in Fig.~\ref{fig_lidar_logo}b. The detection is monostatic and the co-observed backreflection from the collimation lens serves as the zero-distance plane in the measurement.
Target points detected in the back plane are clearly separated due to the cm-level distance precision and accuracy observed on all 30 FMCW channels (cf.~Fig.~\ref{fig_lidar_logo}c,d) and highlighted as filled points.


\subsection*{Discussion \& Conclusion}
\noindent In summary, we have reported a novel method for massively parallel coherent LIDAR utilizing photonic chip-based soliton microcombs. 
It enables to reproduce arbitrary frequency chirps of the narrow linewidth pump laser onto all comb teeth that compose the soliton at speeds beyond 10$^{17}$~Hz$^2$, and has the potential to significantly increase the frame rate of imaging coherent LIDAR systems via parallelization.
In contrast to earlier works in frequency comb-based LIDAR~\cite{Suh2018a,Trocha2018,Kuse2019}, the comb teeth in parallel FMCW LIDAR are spatially dispersed with diffractive optics and separately measure distances and velocities in a \emph{truly parallel} fashion. 
Assuming a similar setup as~\cite{Marin-Palomo2017}, i.e. 179 carriers with 50~GHz spacing in the C+L telecom wavelength bands, we expect aggregate pixel measurement rates of 17.9~Mpixel/s for 100~kHz modulation frequency and 179~Mpixel/s for 1~MHz modulation frequency, well beyond current technologies of long range TOF and FMCW LIDAR systems. 
Although, the slow power modulation of the comb sidebands during the frequency chirp only weakly influences the distance and velocity evaluation, we emphasize that it can be avoided entirely if both the laser and cavity are modulated in unison. Similarly, the laser can be self-injection locked to the modulated cavity, which can furthermore extend the laser coherence length significantly \cite{Liang2015,Pavlov2018}. 
Promising actuation technologies include recently developed high-bandwidth and energy-efficient integrated electro-optical \cite{Zhang2019a} and piezoelectrical actuators \cite{Tian2019}. 

Moreover, by virtue of the laser line separations, our concept is compatible with nanophotonic based gratings for beam separation and could significantly simplify optical phased array systems~\cite{Sun2013}, wherein one axis of beam separation is provided by the nanophotonic grating and a second axis is provided by integrated phase shifters. 
Furthermore, this concept alleviates problems with eye safety, as the light is dispersed over multiple detection pixels at all times, similar to time-of-flight flash systems, yet avoids the problems associated with the excessive peak powers of high-energy pulsed light sources. 
Finally, the approach presented can also be carried out in a dual comb approach, whereby the second comb scans in unison with the first, but exhibits a different repetition rate, which alleviates the need for demultiplexing and individual detection of the comb lines. 
It should be noted that (resonant) electro-optical frequency combs~\cite{Metcalf2013,Zhang2018} based on $\mathrm{LiNbO_3}$ also provide a platform in which the presented approach can be realized.
Hence, we conclude that, combined with concurrent advances in chip-scale lasers, optical beamforming structures, and hybrid electro-optical integration, our approach provides a path towards rapid, precise and simultaneously long-range coherent LIDAR modules suitable for industrial, automotive and airborne applications demanding high-speed 3d imaging in excess of 10M~pixel/s.

\subsection*{Acknowledgments}
\noindent This work was supported by funding from the Swiss National Science Foundation under grant agreement No. 165933. This material is based upon work supported by the Air Force Office of Scientific Research (AFOSR), Air Force Material Command, USAF under Award No. FA9550-15-1-0250. J.R. and W.W. acknowledge support from the EUs H2020 research and innovation program under Marie Sklodowska-Curie IF grant agreement No. 846737 (CoSiLiS) and No. 753749 (SOLISYNTH), respectively. The authors acknowledge useful interactions with Andy Zott from ZEISS AG. Elements of Fig.~\ref{fig_concept}b (tree and car) created by www.freepik.com.

\subsection*{Data Availability Statement}
\noindent All data, figures and analysis code will be published on \texttt{Zenodo} upon publication of the work.

\subsection*{Author contributions}
\noindent A.L. and J.R. conducted the various experiments and analyzed the data. E.L. assisted with laser linearization. W.W. performed the numerical simulations. A.L designed the samples. J.L fabricated the samples in the CMi with the assistance of Rui Wang. All authors discussed the manuscript. J.R., T.J.K., M.K. and E.L. wrote the manuscript. T.J.K. supervised the work and conceived the experiment.

\bibliography{citations_corrected}

\section*{Methods}

\subsection*{Sample details and fabrication}
\noindent Integrated Si$_{3}$N$_{4}$ microresonators are fabricated with the photonic Damascene process~\cite{Pfeiffer2018d}, deep-ultraviolet (DUV) stepper lithography~\cite{Liu2018} and silica preform reflow~\cite{Pfeiffer2018}. 
The waveguide cross-section is 1.5~$\mu$m wide and 0.82~$\mu$m high, with anomalous second order dispersion of $D_{2}/2\pi = 1.13$~MHz and third-order dispersion parameter of $D_{3}/2\pi = 576$~Hz, where the positions of the resonance frequencies close to the pumped resonance are expressed with the series $\omega_{\mu} = \omega_0 + \sum_{i \geq 1} D_i \mu_i / i!$. 
The ring radius is 228.39~$\mu$m and results in a resonator free-spectral-range of $D_{1}/2\pi = 98.9$~GHz, which is chosen to match the 100~GHz telecom ITU grid. 
The resonator is operated in the strongly overcoupled regime with an intrinsic loss rate $\kappa_0/2\pi = 15$~MHz and bus waveguide coupling rate $\kappa_\mathrm{ex}/2\pi = 100$~MHz. 
Operation in the strongly overcoupled regime bears the advantage of suppressing thermal nonlinearities during tuning as well as increasing the power per comb line before and optical signal to noise ratio after post-amplification. 
Input and output coupling of light to and from the photonic chip is facilitated with double inverse tapers~\cite{Liu2018b} and lensed fibres.

\subsection*{FM soliton microcomb generation}
\noindent We set up a frequency-agile pump laser for soliton generation using a CW external cavity diode laser~(ECDL) coupled into an electro-optical phase modulator (EOM) for measurement of the relative laser cavity detuning, and dual Mach-Zehnder modulator (SSB) biased to single sideband modulation, which is driven by a frequency-agile voltage controlled oscillator (VCO, 5-10 GHz) and an arbitrary function generator (AFG). 
The cw laser is amplified to 1.7 W and 1 mW is split off for chirp linearization in a separate imbalanced Mach-Zehnder fibre interferometer for chirp linearization purposes~\cite{Ahn2007,Feneyrou2017a}. 
The dissipative Kerr soliton (DKS) \cite{Leo2010,Herr2014b} is generated by coupling the FM pump laser onto the photonic chip and tuning of the laser into resonance and single soliton state using the established piezo tuning scheme~\cite{Herr2014b,Guo2017}. 
The detuning with respect to the Kerr shifted cavity resonance and the bistable soliton response is monitored using a vector-network-analyzer~(VNA) driving a weak phase modulation via an inline electro-optical-modulator~\cite{Guo2017} and an optical spectrum analyzer (OSA).
The generated soliton is coupled back into optical fibre, the residual pump light is filtered and the soliton pulse train is amplified with a gain-flattening erbium-doped fibre amplifier (EDFA). 
The repetition rate of the soliton pulse train is 99~GHz and the cavity resonance is aligned to the ITU telecom channel C30 at a wavelength of 1553.3~nm using a thermo-electric cooling~(TEC) device located below the active chip. 
While it is possible to directly modulate all comb teeth post DKS generation, this method suffers from excess insertion loss of the SSB and leads to the generation of unwanted RF modulation sidebands at all the comb teeth. Moreover, we want to highlight the feasibility of our scheme irrespective of the choice of laser and microresonator actuation schemes.

\subsection*{Linearization and calibration}
\begin{figure*}[p]
	\includegraphics[width=0.75\linewidth]{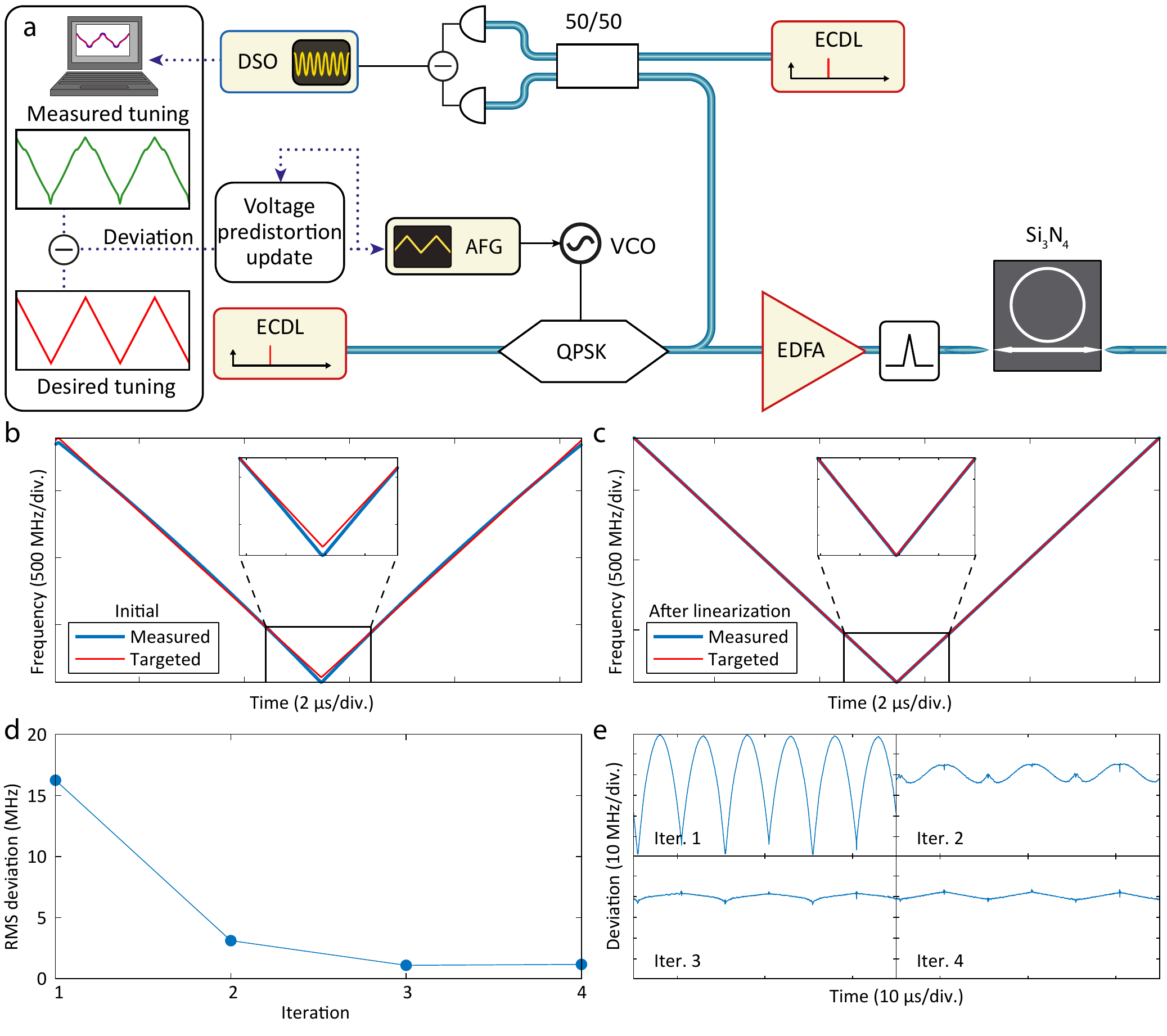}
	\caption{\footnotesize  \linespread{1} \textbf{Pump frequency sweep linearization via the heterodyne method}
		a) Setup for pump laser frequency measurement via heterodyne beat note and chirp linearization feedback.
		b) Initial frequency modulation, when the VCO is driven with a triangular ramp. 
		The measured frequency is compared with the targeted ideal modulation. The ramp frequency is 100~kHz.
		c) Final triangular frequency modulation pattern, after 4 iterations.
		d) Evolution of the RMS frequency deviation during the optimization loop.
		e) Evolution of the deviation between measurement and target sweep, at each iteration of the loop.
		\label{fig_linearize_1}
	}
\end{figure*}

\begin{figure*}[p]
	\includegraphics[width=0.75\linewidth]{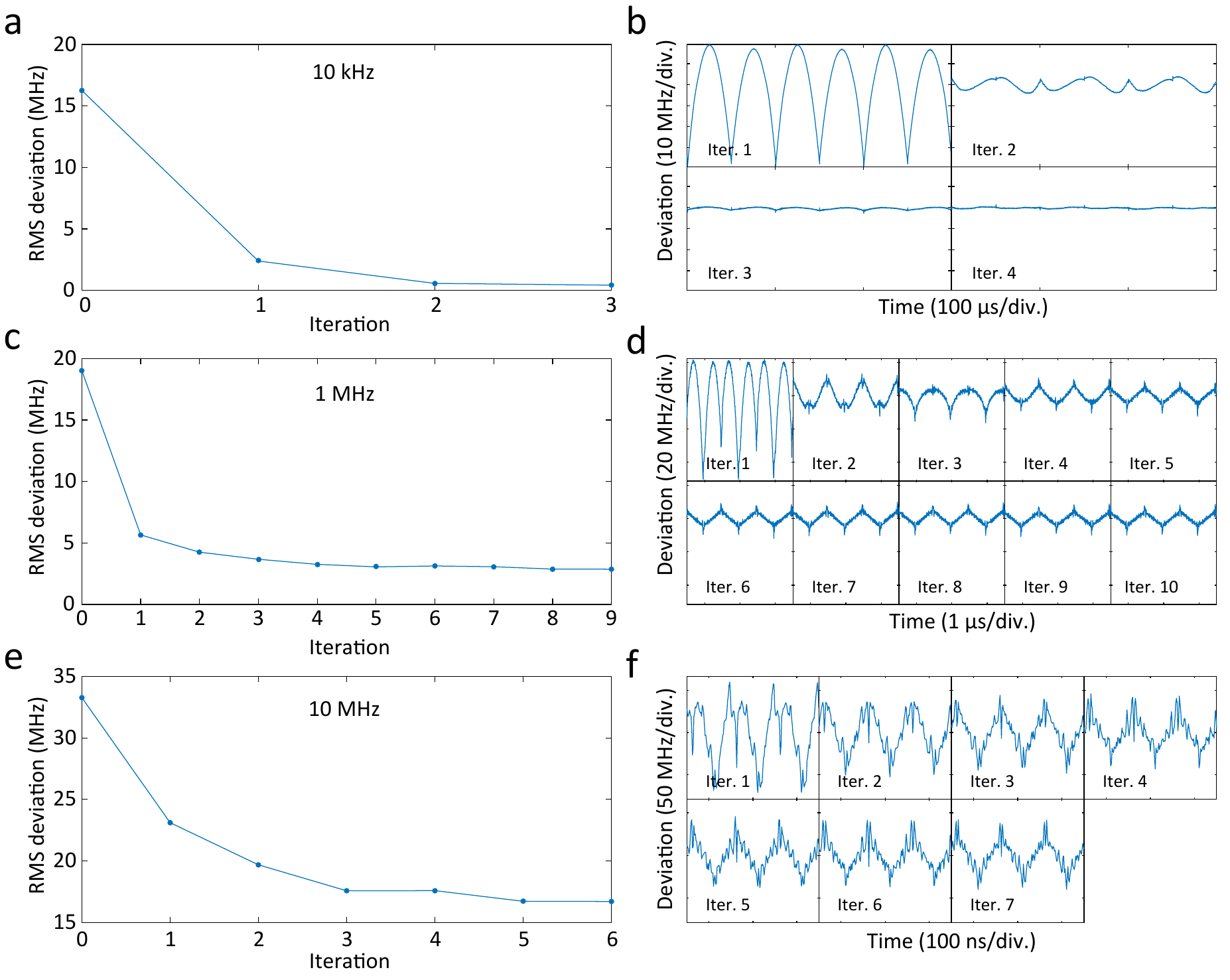}
	\caption{\footnotesize  \linespread{1} \textbf{Linearization results at different modulation frequencies}
		a) c) e) Show the evolution of the RMS frequency deviation during the optimization loop for modulation frequencies of 10~kHz, 1~MHz and 10~MHz, respectively.
		b) d) f) Corresponding evolution of the deviation between the measurement and the target sweep, at each iteration of the loop.
	}
	\label{fig_linearize_2}
\end{figure*}

\begin{figure*}[!p]
	\includegraphics[width=0.75\linewidth]{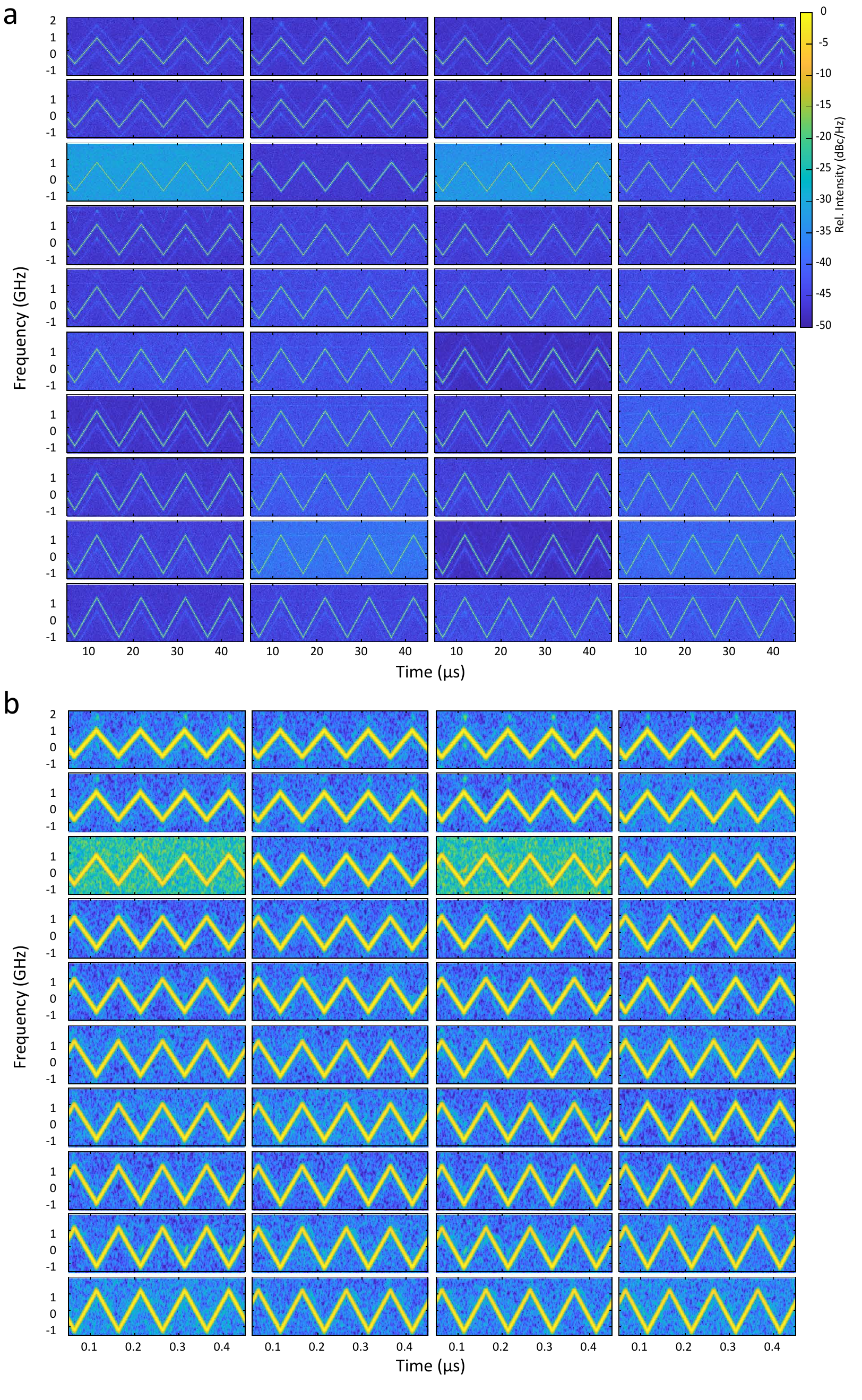}
	\caption{\footnotesize  \linespread{1} \textbf{Channel-by-channel analysis of heterodyne chirp characterization} 
		a)~Time-frequency maps obtained with short-time Fourier transform of the heterodyne beat detection of the individual FMCW channels. 
		Top left to bottom right panels denote optical carriers between 192.1~THz and 196~THz. Modulation frequency 100~kHz. 
		b)~Same as a) but for modulation frequency 10~MHz.
	}
	\label{fig_heterodyne_allchannels}
\end{figure*}

\noindent FM LIDAR requires perfectly linear chirp ramps in order to achieve precise and accurate distance measurements,~\cite{Roos2009}. 
We implemented a digital pre-distortion circuit in order to minimize the chirp nonlinearity of the pump frequency sweep, similar to prior implementations~\cite{Zhang2019c}. 
The optimization procedure was applied in two configurations to measure the pump frequency chirp, either via heterodyne with a reference laser (cf.~Fig.~\ref{fig_chirplinearity_and_linewidth}), or via delayed homodyne detection in an imbalanced Mach-Zehnder interferometer~(MZI). 
The length difference of the calibration MZI arms (12.246~m) is determined using the EOM and VNA and fitting the $sin^2$ spectral response function of the MZI. 
The setup and optimization results for this method are detailed in extended data Fig.~\ref{fig_linearize_1}. 
The chirp is applied to the CW laser with a VCO-driven single sideband modulator. 
The VCO is initially driven by a simple triangular function generated with the AFG. 
The driving voltage is then iteratively corrected to improve the chirp linearity.
After modulation, a fraction of the light is picked up to generate a beat note with a reference external cavity diode laser. 
The downmixed laser frequency is sampled on a real-time oscilloscope~(20~GSa/s) and digitally processed to perform a short-term Fourier transform followed by peak detection. 
The measured frequency evolution is fitted with a perfect triangular function having a fixed target frequency excursion. 
This allows the deviation from this desired frequency chirp to be assessed. 
The frequency deviation is then converted to voltage -- after computing the average voltage-to-frequency coefficient of the VCO -- and then added to the current tuning function of the AFG. 
This procedure effectively addresses the nonlinear response of the VCO, as shown in extended data Fig.~\ref{fig_linearize_1}b-e. 
The optimization procedure was applied successfully at different tuning speed (10~kHz -- 10~MHz), as shown in extended data Fig.~\ref{fig_linearize_2}. 
However, with increasing tuning speed, the residual RMS deviation increases, which we attribute to the limited tuning bandwidth of the VCO.

\subsection*{Heterodyne characterization of FM soliton microcomb}
\begin{figure*}[!p]
	\includegraphics[width=0.85\linewidth]{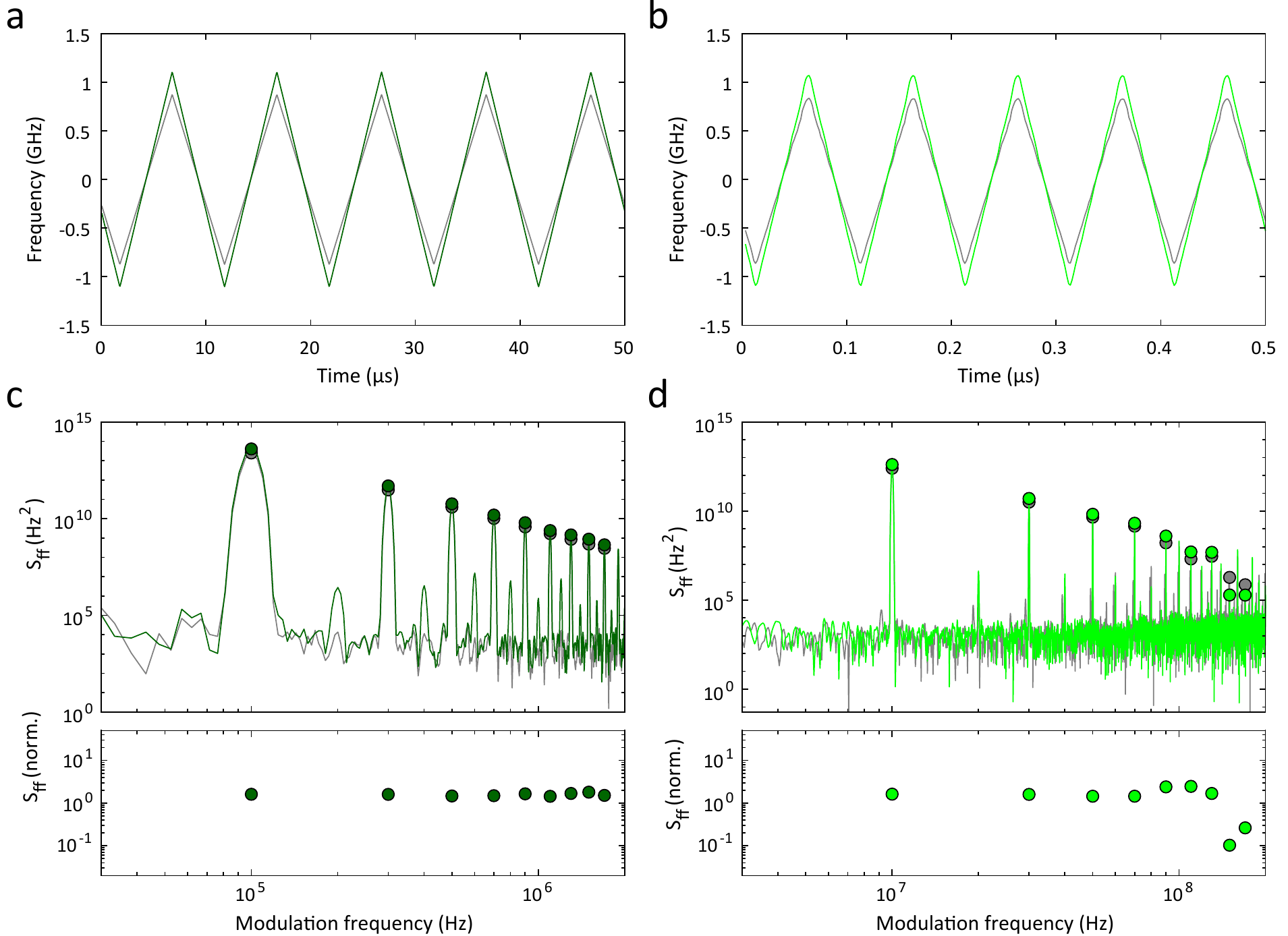}
	\caption{\footnotesize  \linespread{1} \textbf{Frequency-dependent transduction of carrier modulation from pump to comb sidebands} 
		a)~Time dependent frequency of pump laser at 193~THz (grey) and 195~THz comb sideband ($\mu = 20$, dark green) and modulation frequency 100~kHz.
		b)~Same as a), but for modulation frequency 10~MHz.
		c)~Power spectral density of frequency modulation $S_{ff}$ for pump (grey) and sideband (dark green). The Markers denote the positions of harmonics, which are used in the transduction analysis. Bottom: Power spectral density of sideband frequency modulation harmonics normalized to the corresponding modulation power spectral density of the pump (cf.~Fig.\ref{fig_chirplinearity_and_linewidth}).
		d)~Same as c), but for modulation frequency 10~MHz.
	} 
	\label{fig_heterodyne_analysis_method}
\end{figure*}

\noindent Heterodyne characterization of the transduced modulation is carried out to avoid possible ambiguities of delayed homodyne detection and catch high frequency noise components obscured in low bandwidth detection. 
The spectral channels are isolated using a commercial telecom WDM demultiplexer based on planar arrayed waveguide gratings and superimposed on a high-bandwidth (10 GHz) balanced photoreceiver. 
The data is recorded on a high-bandwidth balanced photodetector and a fast realtime sampling oscilloscope. 
Modulation frequencies span from 10 kHz to 10 MHz in our study and are limited by the actuation bandwidth of the AFG and VCO. The total measurement duration is between 0.5~ms (10~kHz) and 30~$\mu$s (10~MHz).
The instantaneous frequency is determined via short-time Fourier transform using a 4th-order Nut-hall window and in case of the pump channel (193~THz) is linearized by applying iterative predistortion of the VCO input (cf.~Fig.~\ref{fig_linearize_1}). The resolution bandwidth $\Delta f$ of the transform window is adjusted to minimize the effective linewidth of the chirped signal. 
\begin{equation}
\Delta f = \sqrt{\dfrac{T}{2B}}
\end{equation}
By tuning the second ECDL close to the individual comb sidebands, we can separately measure the transduced frequency modulation patterns for each comb sideband within the bandwidth of the demultiplexer. 
The resulting time frequency maps for modulation frequencies 100~kHz and 10~MHz across 5~modulation periods are depicted in extended data Fig.~\ref{fig_lidar_allchannels}. 
The tuning nonlinearity of the comb sidebands is calculated as the RMS deviation of the measured tuning curve from a perfect triangular frequency modulation trace determined with least-squares fitting. 
We determine frequency-dependent transduction from the intensities of the \nth{1} to \nth{9} harmonic of the triangular FM spectrum, which we normalize with respect to the corresponding pump modulation amplitude (cf.~extended data Fig.\ref{fig_heterodyne_analysis_method}). 
We observe a slight amplification of the modulation for frequencies around 100 MHz both on the pump and sideband. 
A fine analysis reveals three effects. 
For low modulation frequencies, weak even order sidebands arise, which we attribute to the hysteresis effect, which accompanies the generation of single mode dispersive waves \cite{Yi2017,Guo2017a} essentially introducing a small asymmetry in the transduced chirp.

\subsection*{Parallel velocimetry and ranging}
\noindent The experimental setup is illustrated in Fig.~\ref{fig_lidar_flywheel}a. 
The frequency modulation $1/T$ and excursion $B$ of the microcomb pump are adjusted to 100~kHz and 1.7~GHz, respectively.
The FM comb is amplified with a gain-flattened EDFA and split into signal ($90\%$) and local oscillator ($10\%$) paths. 
A transmission grating~(966~lines/mm) spectrally disperses the individual signal comb lines along the circumference of the flywheel. 
Normal incidence reflection of the wheel is obtained by the FM microcomb sideband at 193.8~THz. 
A bistatic detection with separate collimators for the transmit and receive path is chosen to minimize spurious backreflection in the fibre components.
The back-reflected signal and local oscillator comb lines are spectrally separated in the demultiplexer and superimposed on a balanced photodetector for detection. 
Two 1x40 mechanical optical switches are installed with the demultiplexers to allow individual channels to be measured sequentially, alleviating the requirement to provide 30 balanced photodetectors and analog-to-digital converters. 
We stress that all measurements are done illuminating and receiving light and demultiplexing all the pixels simultaneously. 
Hence any additional noise and crosstalk between the channels would be detected in our setup. 
Yet, our system is impervious to crosstalk and interference between the channels, because of the spectral channel separation, in contrast to simple spatial channel separation~\cite{Martin2018} that requires sequential operation. 
While our current setup utilizes discrete telecom fibre components and optical switches for the detection, we emphasize that high-performance integrated photonic solutions for many-channel DWDM communications have been demonstrated~\cite{Fang2010a} and can be integrated on the Si$_{3}$N$_{4}$ photonic chip with comparable performance as the commercial telecom components employed here~\cite{Ahn2007a,Piels2014}.
The calibration of the channel-dependent frequency excursion bandwidth for the ranging experiments is performed using a second MZI (8.075~m, cf. extended data Fig.~\ref{fig_channel_calibration}). 
The calibration curve is detected once before the start of the measurements and assumed constant throughout. 
The distance and velocity precision and accuracy of the system are determined using a small flywheel (radius~20~mm) mounted on a fast DC motor spinning at up to 228~Hz (cf.~Fig.~\ref{fig_lidar_flywheel}).
The data analysis is performed with simple Fourier transform accounting for a constant 535~ns delay between the AFG and the LIDAR lasers, which is predominantly obtained from the optical fibre lengths of the EDFAs.
Further improvements, especially in long range detection can be achieved using active demodulation analysis \cite{Feneyrou2017}. 

\subsection*{Demonstration of parallel imaging}
\noindent The optical setup is depicted in Fig.~\ref{fig_lidar_logo}~a, wherein the optical receiver, demultiplexers and detectors are omitted for brevity, but are set up as depicted in Fig.~\ref{fig_lidar_flywheel}a. 
The target is composed of two sheets of white paper spaced by 11.5~cm. 
The EPFL university logo, (width~7.5~cm, height~22.7~cm) is cut from the first sheet and oriented vertically. 
The FMCW LIDAR channels are dispersed horizontally using a 966 lines/mm transmission grating and directed to the target with a 45$^{\circ}$ steering mirror.
A monostatic detection scheme using an optical circulator and single collimator Fig.~\ref{fig_concept} is chosen. The detector aperture is increased by placing a 75 cm focal length lens 1~m away from the 4 mm collimator and behind the grating.
We note that modal interactions with the fundamental TM mode strongly increase the power fluctuation of channels at 195.2 and 195.3 THz and spoils their use in FM LIDAR experiments by shortening the effective sampling length.

\section*{Extended data figures}
\renewcommand{\figurename}{\textbf{Extended Data Fig.}}

\begin{figure*}[!p]
	\includegraphics[width=0.75\linewidth]{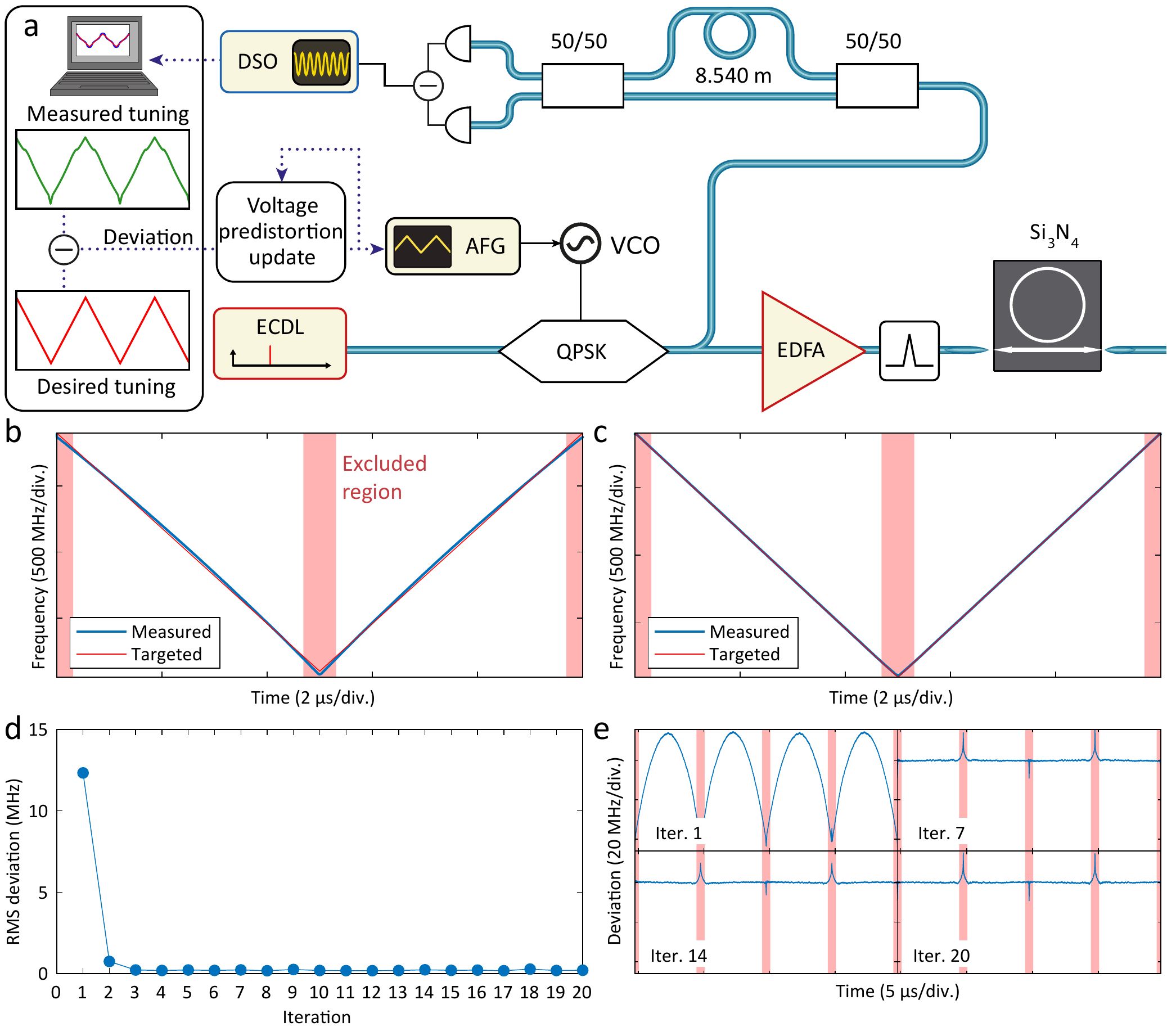}
	\caption{\footnotesize  \linespread{1} \textbf{Channel-by-channel analysis of delayed homodyne chirp characterization} 
		a)~Time-frequency maps obtained with short-time Fourier transform of the heterodyne beat detection of the individual FMCW channels. 
		Top left to bottom right panels denote optical carriers between 192.1~THz and 196~THz. Modulation frequency 100~kHz. 
		b)~Same as a) but for modulation frequency~10 MHz.
	} 
	\label{fig_linearize_3}
\end{figure*} 

\begin{figure*}[!p]
	\includegraphics[width=0.7\linewidth]{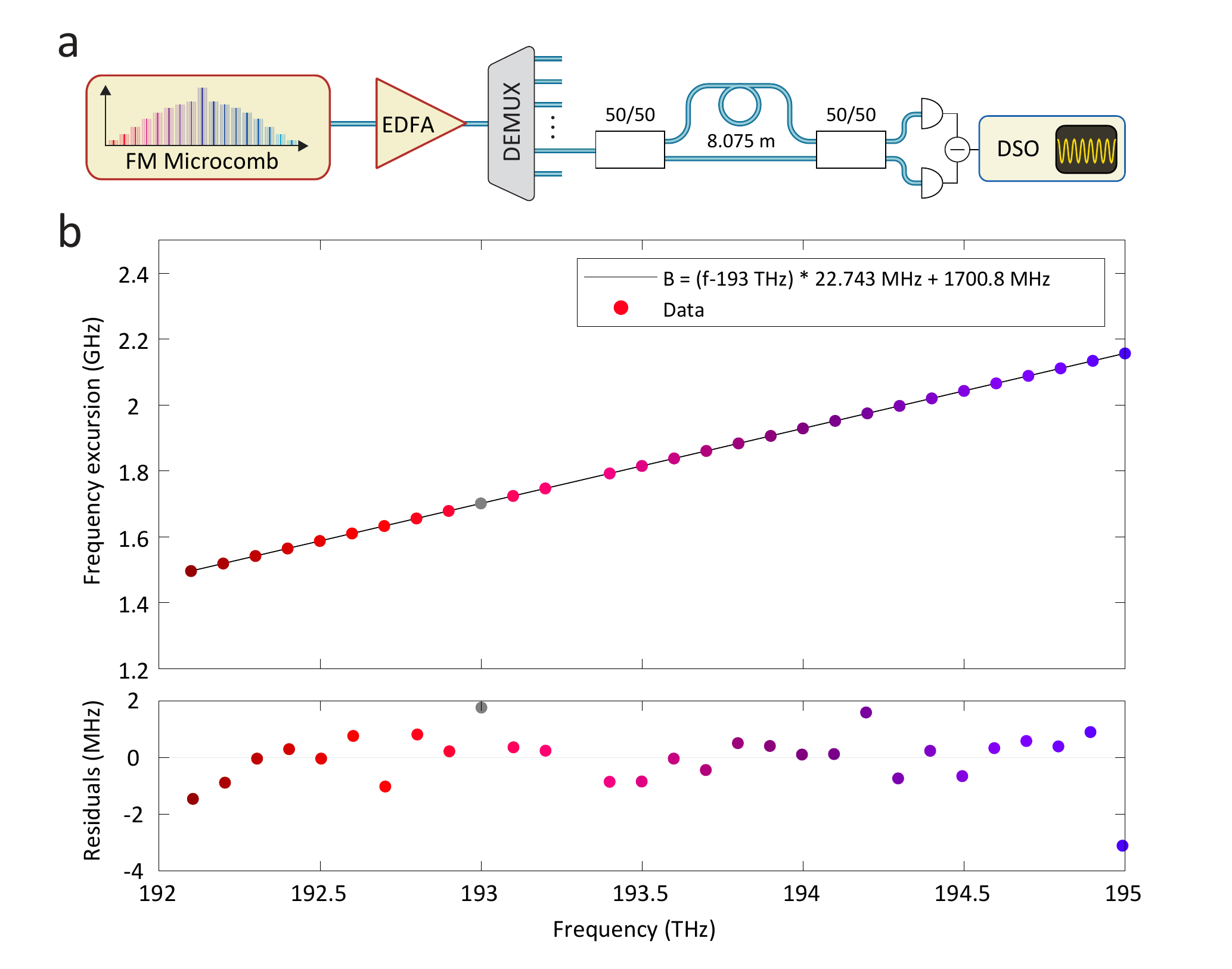}
	\caption{\footnotesize  \linespread{1} \textbf{Calibration of channel dependent frequency excursion bandwidth for distance and velocity measurements} 
		a)~Measurement setup. Linearized FM microcomb (cf. extended data Fig.\ref{fig_linearize_3} for setup) is amplified and individual channels are selected by connecting the LO path of the measurement setup to a calibrated imbalanced MZI (8.075~m). 
		b)~Top:~Frequency-excursion bandwidth determined from measurement of independently measured length of imbalanced MZI. Linear fit related to Raman self-frequency shift. Bottom:~Residuals of linear fit.
	} 
	\label{fig_channel_calibration}
\end{figure*}

\begin{figure*}[!p]
	\includegraphics[width=0.75\linewidth]{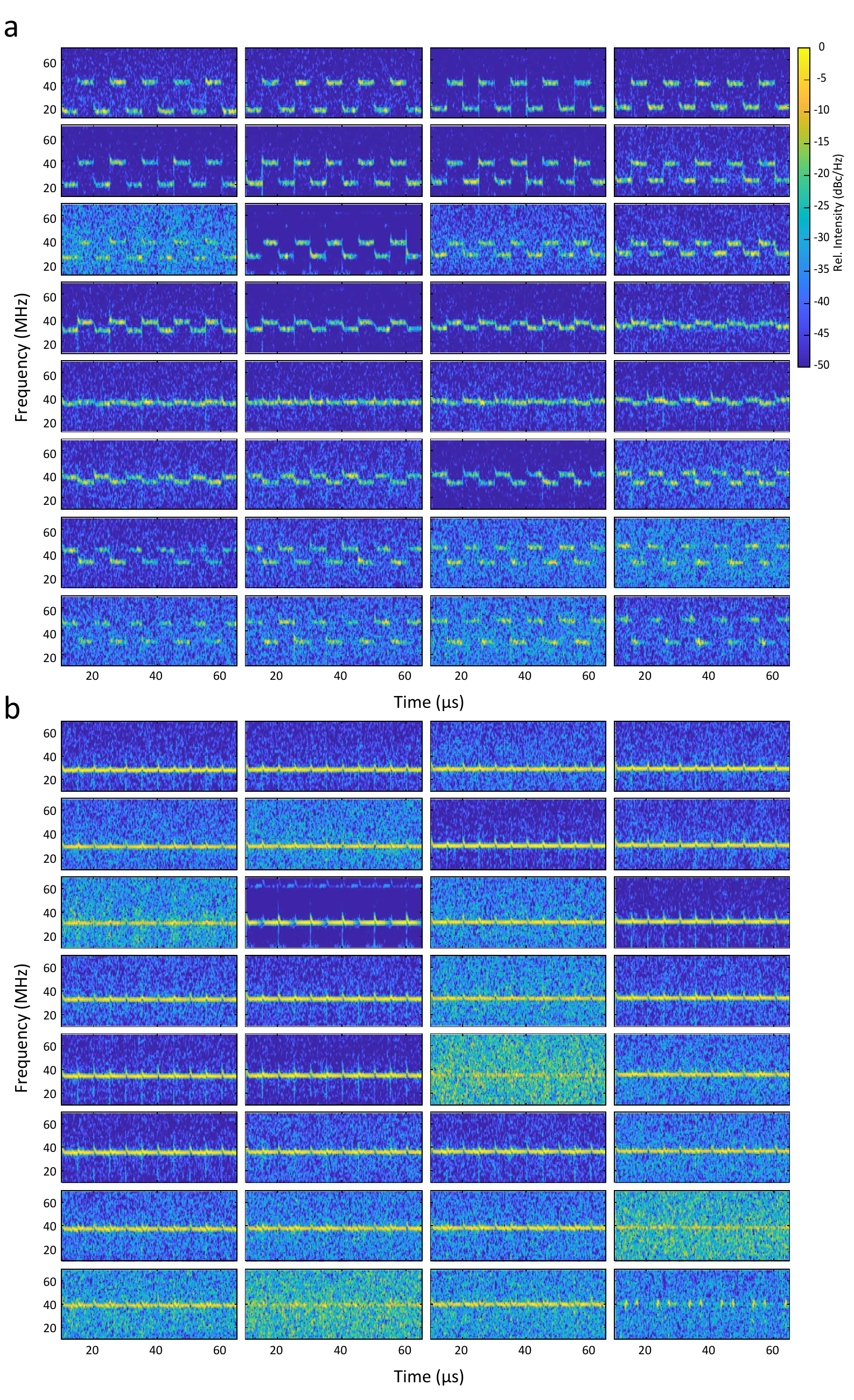}
	\caption{\footnotesize  \linespread{1} \textbf{Channel-by-channel analysis of proof-of-concept LIDAR demonstration} 
		a)~Time-frequency maps obtained with short-time Fourier transform of the delayed homodyne beat detection of the individual FMCW channels back-reflected from the rotating flywheel. Top left to bottom right panels denote optical carriers between 192.1~THz and 195.2~THz. Modulation frequency 100~kHz. 
		b)~Same as a) but for static flywheel.
	} 
	\label{fig_lidar_allchannels}
\end{figure*} 

\end{document}